# Superconductivity at 250 K in lanthanum hydride under high pressures

A. P. Drozdov[1], P. P. Kong[1], V. S. Minkov[1], S. P. Besedin[1], M. A. Kuzovnikov[1,6], S. Mozaffari[2], L. Balicas[2], F. Balakirev[3], D. Graf[2], V. B. Prakapenka[4], E. Greenberg[4], D. A. Knyazev[1], M. Tkacz[5], and M. I. Eremets[1]

[1]*Max-Planck-Institut fur Chemie, Hahn-Meitner Weg 1, 55128 Mainz, Germany*
[2]*National High Magnetic Field Laboratory (NHMFL), Florida State University, Tallahassee, Florida 32310, USA*
[3]*NHMFL, Los Alamos National Laboratory, MS E536, Los Alamos, New Mexico 87545, USA*
[4]*Center for Advanced Radiation Sources, University of Chicago, 5640 South Ellis Avenue, Chicago, Illinois, 60637, USA*
[5]*Institute of Physical Chemistry PAS, Kasprzaka 44/ 52, 01-224 Warsaw, Poland*
[6]*Institute of Solid State Physics RAS, Chernogolovka, Moscow District, 142432 Russia*

**The discovery of superconductivity at 203 K in $H_3S$[1] brought attention back to conventional superconductors whose properties can be described by the Bardeen-Cooper-Schrieffer (BCS) and the Migdal-Eliashberg theories. These theories predict that high, and even room temperature superconductivity (RTSC) is possible in metals possessing certain favorable parameters such as lattice vibrations at high frequencies. However, these general theories do not suffice to predict real superconductors. New superconducting materials can be predicted now with the aid of first principles calculations based on Density Functional Theory (DFT). In particular, the calculations suggested a new family of hydrides possessing a clathrate structure, where the host atom (Ca, Y, La) is at the center of the cage formed by hydrogen atoms[2-4]. For $LaH_{10}$ and $YH_{10}$ superconductivity, with critical temperatures $T_c$ ranging between 240 and 320 K is predicted at megabar pressures[3-6]. Here, we report superconductivity with a record $T_c \sim 250$ K within the $Fm3m$ structure of $LaH_{10}$ at a pressure $P \sim 170$ GPa. We proved the existence of superconductivity at 250 K through the observation of zero-resistance, isotope effect, and the decrease of $T_c$ under an external magnetic field, which suggests an upper critical magnetic field of ~120 T at zero-temperature. The pressure dependence of the transition temperatures $T_c(P)$ has a maximum of 250-252 K at the pressure of about 170 GPa. This leap, by ~ 50 K, from the previous $T_c$ record of 203 K[1] indicates the real possibility of achieving RTSC (that is at 273 K) in the near future at high pressures and the perspective of conventional superconductivity at ambient pressure.**

The quest for room temperature superconductivity is a longstanding challenge. Superconductivity was considered as a low-temperature phenomenon as the known materials had $T_c$s inferior to ~30 K, until 1986, when Bednorz and Müller discovered the cuprates[7] – copper based superconductors with enormous $T_c$s that can reach 164 K[8] – or the so-called high temperature superconductors (HTSCs). The cuprates stimulated an intense quest for room temperature superconductivity. However, the maximum value of $T_c$ has remained at the same for 25 years despite tremendous efforts.

The discovery of superconductivity at 203 K in $H_3S$[1] at high pressures offered another route in the search for high temperature superconductivity – in conventional superconductors[9]. In fact, that was the first confirmation of the old implication of the BCS and Migdal-Eliashberg theories concerning the possibility of high temperature superconductivity in materials with high phonon frequencies[10]. Hydrogen- and carbon-abundant materials can, in principle, provide the required high frequencies in the phonon spectrum as well as the strong electron-phonon interaction[11,12]. A great support for narrowing down the experimental search for RTSC came from crystal structure predictions based on density functional theory[13-16]: as soon as the structure is determined, the electron and phonon spectra as well as the transition temperatures can be estimated from the BCS and Eliashberg theories. First principle theory of superconductivity is developing too[17,18]. Recently, nearly all of the binary hydrides have been studied theoretically[13-15], with the calculations now focusing on the ternary compounds[19].

Through this broad theoretical search for RTSC, an interesting family of hydrides having a clathrate structure was found, leading to a first report on $CaH_6$[2] and subsequently on $YH_6$[20]. Here Ca and Y are located at the center of the $H_{24}$ cages, and act as electron donors contributing to electron pairing, while the H atoms are weakly covalently bonded to each other within the cage. Their structure is quite different from the one of $H_3S$ where each hydrogen atom is strongly, covalently connected to the two nearby sulfur atoms. The clathrate structure with even larger hydrogen content, $H_{32}$ cages, were later predicted in $YH_{10}$ and in rare earth (RE) hydrides[3] such as $LaH_{10}$[3,4,21]. These superhydrides can be considered to be doped versions of metallic hydrogen and therefore are naturally expected to have high $T_c$s. Indeed, DFT predicts a $T_c$ of 235 K at 150 GPa for $CaH_6$[2], $T_c$ = 305–326 K at 250 GPa[5] (or 303 K at 400 GPa[4]) for $YH_{10}$, and a $T_c$ ~ 280 K at ~200 GPa for $LaH_{10}$[3,4].

Obviously, these remarkable predictions for nearly room temperature superconductivity motivate experimental verifications, although the experiments are very challenging. The first lanthanum superhydride was synthesized only recently under $P$>160 GPa upon heating up to ~1000 K[22]. The X-ray data indicate that the stoichiometry corresponds to $LaH_{10\pm x}$ (-1 < x < 2) which is close to the predicted $LaH_{10}$[3,5]. Somayazulu *et al.* measured the temperature dependence of the resistivity of La heated with $NH_3BH_3$ as the hydrogen source under similar pressure and observed a drop in the resistance at ~260 K upon cooling and ~248 K upon warming the sample which was assigned to the superconducting transition of $LaH_{10\pm x}$[23]. They also observed a series of resistance anomalies at temperatures as high as 280K. However, neither a zero resistance state nor additional confirmations (like the Meissner effect or isotopic effects or the effect of an external magnetic field on the transition temperature) was provided. Simultaneously, a record superconducting transition of $T_c$ = 215 K in $LaH_x$ displaying zero resistance after the transition was reported[24].

In the present work we performed an extensive search for superconductivity in the lanthanum-hydrogen system. In our experiments (see Methods for details) we found a number of superconducting transitions at $T_c$ ~250 K, 215 K, 110 K, and 70 K (Fig. 1 and Extended Data, Figs 1-5) with a sharp drop in the electrical resistance towards zero value.

To confirm the existence of superconductivity it is crucial to detect the Meissner effect, but magnetization measurements within the diamond anvil cell (DAC) in a SQUID magnetometer are problematic due to the small sample volume. The Meissner effect was detected in the case of $H_3S$[1] but on a sample having a diameter of ~100 µm[1]. Our typical lanthanum hydride samples are much smaller (10-20 µm), therefore, their magnetization signal is below the sensitivity of a SQUID magnetometer. These small samples are challenging for the AC magnetic susceptibility method too and hence it will require further experimental developments.

Nevertheless, the superconducting nature of the transition can be verified *via* its dependence on the external magnetic field, since the magnetic field reduces $T_c$ in type II superconductors through the so-called orbital effect or by breaking the spin-singlet state of the Cooper pair. Figure 2 shows that an applied field of $\mu_0H$= 9 T indeed reduces the onset of the superconducting transition by about 10 K. The extrapolation of the temperature dependent upper critical fields $H_{c2}(T)$ towards $T$ = 0 K, Fig. 2b, yields values between 95 and 136 T for $H_{c2}(0)$. Notice the two steps near 245 K and 230 K in the superconducting transition at zero-field. The higher temperature step gradually broadens with increasing magnetic field and completely disappears above 3 T. This behavior is consistent with inhomogeneous superconductivity. While it is difficult to investigate the local inhomogeneity of the superconducting state in a DAC, multiple examples of multi-step transitions in inhomogeneous samples at ambient pressure can be found in the literature[25]. In particular, an anomaly resembling a double transition, as well as resistance peaks at $T_c$, were also observed in superconducting boron-doped diamond, where it was ascribed to the extinction of individual superconducting quasiparticles prior to the onset of global phase-coherence in strongly inhomogeneous samples[25]. We note that some degree of inhomogeneity is inevitable in samples synthesized within the confined space provided by the DAC.

Further study of the superconductivity in $LaH_{10}$ includes the pressure dependence of the superconductivity to determine the highest $T_c$. It clearly has a "dome"-like shape: after the initial increase and reaching the maximum $T_c$ = 250-252 K at ~170 GPa, $T_c$ decreases abruptly at higher pressures (Fig. 1, inset). This is in a clear disagreement with the claims of superconductivity with $T_c$ =280 K and 290 K in Ref.[23].

The structure of the superconducting phase with $T_c$ ~ 250 K at 150 GPa, was determined as face-centered cubic (*fcc*) lattice *Fm3m* with the refined lattice constant $a$ = 5.1019(5) Å (V = 132.80(4) Å$^3$) (Fig. 3c). This *fcc* structure is in agreement with the previous experimental study[22] and theoretical predictions for $LaH_{10}$[3-5] with volumes of 131.9 Å$^3$ at 172 GPa and 128.8 Å$^3$ at 175 GPa. Experimentally, the stoichiometry can be estimated from the volume occupied by H atoms: ~18.2 A$^3$ per La atom – the difference between this volume and the volume occupied by the lanthanum atom[22]. The volume occupied by the hydrogen atom(s) was accurately determined from the diffraction study of $LaH_3$ (Extended Data, Fig. ED2): ~1.9 Å$^3$ at 152 GPa. This gives the stoichiometry $LaH_{9.6}$ for the 250 K superconductive phase, in good agreement with the predictions for $LaH_{10}$[3,5].

The isotope effect, i.e. the shift in $T_c$ through replacement of hydrogen by deuterium, gives direct evidence on the pairing mechanism of superconductivity[9]. For that, $T_c$s in hydrides should be compared with those in deuterides in the same phases and under the same pressures.

A complication arose from the existence of competing phases: in hydrides different phases with $T_c$ =250 K (*Fm3m* structure) and 215 K (the structure to be determined) are formed under the same pressure-synthesis conditions. This assertion is valid also for the synthesized deuteride phases displaying $T_c$ of ~168 K and ~140 K. The later phase has tetragonal *P4/nmm* structure and $LaD_{11}$ stoichiometry (Fig. 3e and Extended Data, Fig. 5) in agreement with the predicted $LaH_{11}$ phase[3]. The computational calculations[3] suggest that the tetragonal phase of $LaH_{11}$ is more stable than the close packed cubic phase of $LaH_{10}$. Note, that the calculations predict very low $T_c$ for the $LaH_{11}$ phase[3] contrary to our observations.

At present we cannot directly compare $T_c$s in hydrides and deuterides having the same crystallographic structures. We can naturally suppose that the hydride with the maximum $T_c$ (250 K) and the deuteride with maximum $T_c$ (168 K) have the same crystal structure, that is the *Fm3m* phase (see Extended Data for details). Similarly, the hydride with $T_c$ = 215 K and the deuteride with $T_c$ = 140 K should have a common *P4/nmm* structure. Under this assumption, we obtained the isotope coefficient $\alpha$ of about 0.57 and 0.62 for the superconductivity in the *fcc* and *P4/nmm* phases, respectively. These values of $\alpha$ are close to 0.5, which is the maximum expected value for a phonon-mediated superconductivity in a harmonic approximation. The values of $\alpha$>0.5 could be explained probably by a soft anharmonism of H/D vibrations ($\omega_H/\omega_D < \sqrt{2}$, where $\omega_H$ and $\omega_D$ are phonon frequencies for hydrogen and deuterium, respectively). The isotope coefficient was determined from the BCS equation $T_c = A \cdot m^{-\alpha}$; where $m$ is the isotope mass and A is a constant. These values of $\alpha$ are close but higher than the classical value of $\alpha \approx 0.5$ for conventional superconductivity. The work is in progress to determine the crystal structure of the hydride with $T_c$=215 K and deuteride with $T_c$~168 K.

We found a record $T_c$ = 250 K for $LaH_{10}$ belonging to the lanthanum – hydrogen system, thus confirming the prediction of high temperature superconductivity in superhydrides with the sodalite-like clathrate structure first proposed for $CaH_6$[2]. Our study makes a leap forward on the road to the room-temperature superconductivity, and also provides an evidence that the art-of-state methods for crystal structure prediction can be very useful for the search of high temperature superconductors. The current theoretical predictions for RTSC in yttrium superhydrides[3,5] motivate further experiments. We hope that the observation of superconductivity with very high $T_c$(s) found at high pressures will generate interest for further theoretical and experimental efforts concentrated on the search for high temperature conventional superconductors at ambient pressure, for example, in carbon-based materials[11]. One encouraging example is the discovery of superconductivity with $T_c \cong 55$ K in Q-carbon[26].

**Figure captions**

**Fig. 1| Observation of superconducting in $LaH_{10}$.** Superconducting transitions in lanthanum superhydride $LaH_{10}$ measured in different samples synthesized from a La+$H_2$ mixture: red curve corresponds to the sample heated up under 145 GPa displaying $T_c$ of ~244 K, which shifts to ~249 K when the pressure is increased up to 151 GPa (orange curve); dark yellow curve corresponds to the sample heated under 135 GPa with a $T_c$ of ~245 K; blue curve corresponds to a sample heated under 150 GPa with $T_c$ ~249 K. Red, orange and dark yellow curves show the sharpest transitions to zero-resistance upon cooling. Blue curve, as well as many others samples, shows onsets of the superconductive transition around the same temperatures but the sharp superconducting step being distorted by the presence of an impurity phase and/or inhomogeneity in the sample. The resistance of the samples was divided by the shown coefficients for the sake of clarity. A vertical line drawn at 273 K marks the RTSC limit. Inset: pressure dependence of $T_c$ for the 6 different samples.

**Fig. 2| Superconducting transition under an external magnetic field.**
**a**, Electrical resistance $R$ as a function of the temperature $T$ for $LaH_{10}$ under applied magnetic fields up to $\mu_0 H = 9$ T. The width of the superconducting transition remains essentially constant up to 9 T. Both, the cool down and the warm up temperature sweeps are plotted. The superconducting critical temperatures were determined as the average between both sweeps. An applied field of $\mu_0 H = 9$ T reduces the onset of the superconducting transition from ~250 K (as extracted from the warm up curve) to ~240 K. Notice that the step observed in the superconducting transition measured at zero-field, which appears around 245 K, disappears under the application of a modest field of just $\mu_0 H = 3$ T. The gradual disappearance of this step via the application of an external magnetic field is shown in the inset. Likely, this step results from inhomogeneities in the superconducting sample. Similar behavior was found in disordered superconducting diamond doped with boron.[25]
**b**, Upper critical fields $H_{c2}$ as a function of the temperature following the criteria of 90% (solid markers) and 50% (open markers) of the resistance in the normal state. Solid curves are extrapolations resulting from fits to the Ginzburg-Landau expression in order to estimate the upper critical fields in the limit of zero temperatures. The temperature dependence of the observed upper critical fields $H_{c2}(T)$ were obtained from the $R(T, H)$ data displayed in Fig. 2(a). The $H_{c2}$s near $T_c$ increase nearly linearly with $T$ as $T$ is lowered. Here, we use a simple Ginzburg-Landau expression $H_{c2}(T) = H_{c2}(0)\left(1 - \left(T/T_c\right)^2\right)$ to estimate the upper critical fields $H_{c2}(0)$ in the limit of zero temperatures. The Ginzburg-Landau coherence lengths deduced from these $H_{c2}(0)$ values ranges between $\xi = 1.56$ and 1.86 nm.

**Fig. 3| Structural analysis.**
X-ray diffraction studies of superconducting lanthanum hydrides. (a) Cake representation of the typical X-ray powder diffraction pattern measured for the sample showing a superconducting step at ~249 K under 150 GPa (blue curve in Figure 1). The dominant face-centred $Fm$-$3m$ phase gives a spotty powder pattern. Integrated powder patterns for samples synthesized from the La+$H_2$ mixture exhibiting a $T_c$ of ~249 K under 150 GPa (b) and the La+$D_2$ mixture with a $T_c$ ~140 K at 142 GPa (c). Indexation and refinement of these powder diffraction patterns yield two different phases, i.e. $Fm$-$3m$ ($a$ = 5.1019(5) Å, $V$ = 132.80(4) Å$^3$) and $P4/nmm$ ($a$ = 3.7258(6) Å, $c$ = 5.0953(12) Å and V =

70.73(2) Å$^3$) with stoichiometries corresponding to LaH$_{9.6}$ and LaD$_{10.6}$, respectively (see text and Extended Data for details). Both refined crystal structure models are in good agreement with the predicted structures for LaH$_{10}$ (Ref. 2) and LaH$_{11}$ (Ref. 3), except for the reflection at ~6.10° that cannot be indexed to the *P4/nmm* structural model (possibly it comes from an impurity phase). Black, red and blue graphs correspond to experimental data, fitted data, and the difference between experimental and fitted data. Le Bail method was chosen because it is impossible to extract the real intensities of the reflections from the spotty patterns. The distribution of these *Fm-3m* (d) and *P4/nmm* (e) phases throughout the samples is very smooth, containing more than 80% of the sample.

**Author contributions**
A.D., P.K., V.M., S.B., M.K., M.T., and D.K. performed experiments on preparation of the samples and measuring the superconducting transition, S.M., D.G., F.B. and L.B. performed studies under external magnetic fields. V.P., E.G., V.M., and M.K. performed X-ray diffraction studies. M.E., V.M, and S.M. wrote the manuscript, with input from all co-authors.
A.D., P.K., V.M., S.B., and M.K. contribute equally. M.E. guided the work.

**Acknowledgements.** M.E. is thankful to the Max Planck community for the invaluable support, and U. Pöschl for the constant encouragement. L.B. is supported by DOE-BES through award DE-SC0002613. S.M. acknowledges support from the FSU Provost Postdoctoral Fellowship Program. The NHMFL acknowledges support from the U.S. NSF Cooperative Grant No. DMR-1644779, and the State of Florida. Portions of this work were performed at GeoSoilEnviroCARS (The University of Chicago, Sector 13), Advanced Photon Source (APS), Argonne National Laboratory. GeoSoilEnviroCARS is supported by the National Science Foundation - Earth Sciences (EAR - 1634415) and Department of Energy- GeoSciences (DE-FG02-94ER14466). This research used resources of the Advanced Photon Source, a U.S. Department of Energy (DOE) Office of Science User Facility operated for the DOE Office of Science by Argonne National Laboratory under Contract No. DE-AC02-06CH11357.

**Extended data** is available for this paper at https
**Correspondence and requests for materials** should be addressed to M.E.

**Methods**

We typically synthesized lanthanum hydride *via* a direct reaction of lanthanum (Alfa Aesar 99.9%) and hydrogen (99.999%) at high pressures. For that, a piece of La was placed into the hole, which was drilled in an insulating gasket. Sample handling was done in an inert $N_2$ atmosphere in a glovebox with $O_2$ and $H_2O$ residual contents of <0.1 ppm. The DAC was preliminary clamped inside the glovebox and transferred into a hydrogen gas loader, opened there under a hydrogen atmosphere. The hydrogen gas was then pressurized up to ~ 0.1 GPa, and the DAC was clamped, the pressure was typically increased to ~ 2 GPa during the clamping. After that, the DAC was extracted from the gas loader, and the pressure was further increased to a desirable value of ~ 120-190 GPa.

Heating of the sample with a laser leads to the formation of a variety of hydrides (Extended Data Figs 1-5). Formation of a particular hydride depends on the pressure, heating temperature, and the amount of hydrogen surrounding the sample. Hydrides containing a large amount of hydrogen (superhydrides) (Extended Data Figs. 3-5) were synthesized only under an evident excess of hydrogen. Under a hydrogen deficient environment, $LaH_3$ (Extended Data Fig. 1) or a variety of different phases (Extended Data Figs. 1,2,5) can form.

For the thermal treatment, one-sided pulsed radiation from a YAG laser was focused onto a spot having ~10 µm in diameter. The sample can be heated up to ~1500 K but typically the temperature remained below ~700 K as we did not observe glowing. This gentle heating was enough to initiate the reaction as the sample expanded and reflected light from the spot where the laser was focused. The laser spot was scanned over the sample to perform a chemical reaction as complete and uniform as possible. Some samples were prepared not from elemental lanthanum as a starting material but from $LaH_3$ which was prepared and analysed earlier[1]. One of the advantages of this method is to have higher hydrogen content. Finally, we found that the superhydride with $T_c$=250 K can be synthesized without any laser heating by just keeping a clamped mixture of $LaH_3$ and hydrogen under a high pressure of ~140 GPa at room temperature for about two weeks.

To verify if the superconductivity is conventional, i.e. dependent upon the phonon frequencies, we substituted hydrogen with deuterium (99.75% purity).

The necessity to perform electrical measurements for a direct proof of superconductivity brings other complications. Typically, the diamonds used in DACs had a culet with a diameter of 40-50 µm and were bevelled at 8° to a diameter of about 250 µm. Tantalum electrodes were sputtered onto the surface of one of the diamond anvils in the van der Pauw four probe geometry. Total resistance of the current leads was typically about 100 Ohm per electrode. A four probe measurement scheme was essential to separate the sample signal from the parasitic resistance of the current leads. A metallic gasket (T301 steel) should be electrically separated from the electrodes by an insulating layer. We prepared this layer from a mixture of epoxy and $CaF_2$ (or a number of other similar materials, e.g. MgO, $CaSO_4$, cBN). The most difficult aspect is to ensure electrical contact between the sample and the electrodes. For that the sample should be large enough to be squeezed between the anvils and pressed against the electrodes, but also small enough to accommodate an excess of surrounding hydrogen. Typical sizes for these samples were 5-10 µm. The laser heating of the pressurized samples is another experimental challenge due of the high thermal conductivity and comparatively high volume of the diamond anvils.

After the synthesis, the temperature dependence of the electrical resistance was measured upon cooling and warming of the samples. One observes hysteresis through this cycling due to the thermal mass of the pressure cell. We present resistance measurements upon warming the DAC as it yields a more accurate temperature reading: the cell was warmed up slowly (~ 0.2 K min$^{-1}$) under an isothermal environment (no coolant flow). Temperature was measured by a Si diode thermometer attached to the DAC with an accuracy of about 0.1 K. All electrical measurements were done at the current set in a range of $10^{-5}$-$10^{-3}$ A, no apparent effect of a current value on the $T_c$ was observed.

Pressure was measured from the frequency of the vibron of the hydrogen surrounding the sample[2], and from the edge of the Raman signal from diamond pressure scale[3]. Typically, the pressure values determined from the $H_2$ ($D_2$) vibron was lower than the ones determined from the diamond Raman edge scale by 10-30 GPa at the pressures of about 150 GPa. We used the hydrogen (deuterium) scale throughout the manuscript (except in cases where the hydrogen vibron could not be observed).

Altogether, we performed dozens of experiments. We used three types of DACs. In particular, DACs with diameters of 20 mm and 8.8 mm were made of nonmagnetic materials, allowing measurements under magnetic fields at the NHMFL in Tallahassee. The X-ray diffraction measurements were done with wavelengths 0.3344 Å and 0.2952 Å, x-ray spot size ~3x4 µm, and Pilatus 1M CdTe detector at the beamline 13-IDD at GSECARS, Advanced Photon Source, Argonne National Laboratory (Chicago). Primary processing and integration of the powder patterns were made using the Dioptas software[4]. The Indexing and refinement were done with GSAS and EXPGUI packages[5].

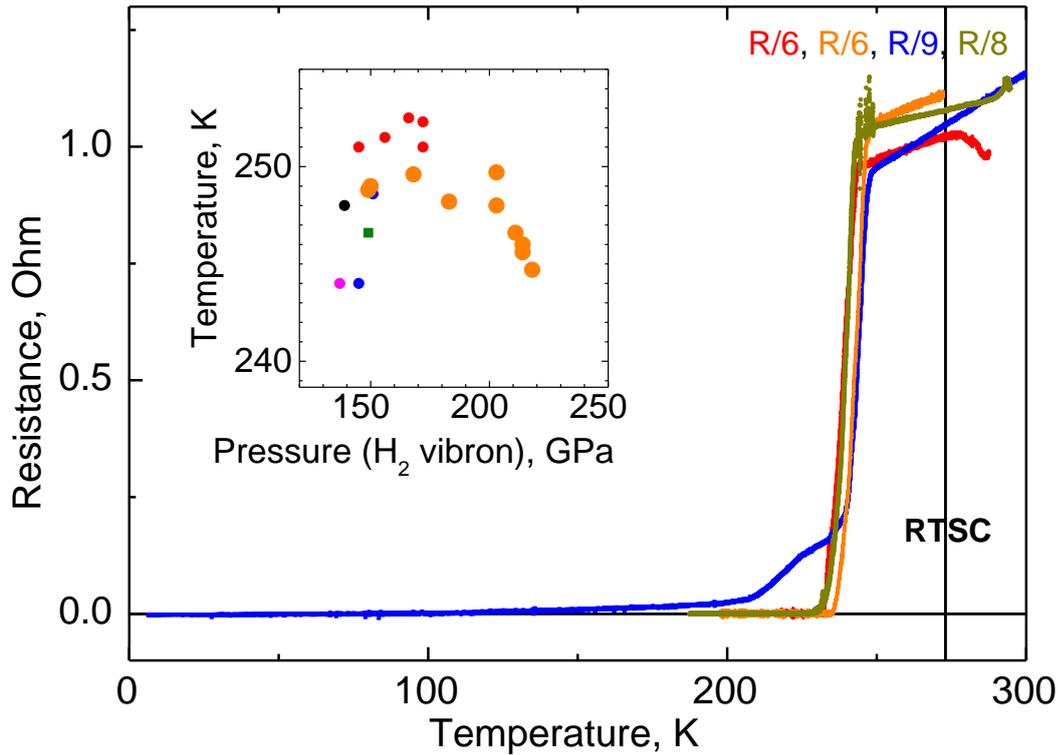

**Fig. 1. Observation of superconducting in LaH$_{10}$.** Superconducting transitions in lanthanum superhydride LaH$_{10}$ measured in different samples synthesized from a La+H$_2$ mixture: red curve corresponds to the sample heated up under 145 GPa displaying $T_c$ of ~244 K, which shifts to ~249 K when the pressure is increased up to 151 GPa (orange curve); dark yellow curve corresponds to the sample heated under 135 GPa with a $T_c$ of ~245 K; blue curve corresponds to a sample heated under 150 GPa with $T_c$ ~249 K. Red, orange and dark yellow curves show the sharpest transitions to zero-resistance upon cooling. Blue curve, as well as many others samples, shows onsets of the superconductive transition around the same temperatures but the sharp superconducting step being distorted by the presence of an impurity phase and/or inhomogeneity in the sample. The resistance of the samples was divided by the shown coefficients for the sake of clarity. A vertical line drawn at 273 K marks the RTSC limit. Inset: pressure dependence of $T_c$ for the 6 different samples.

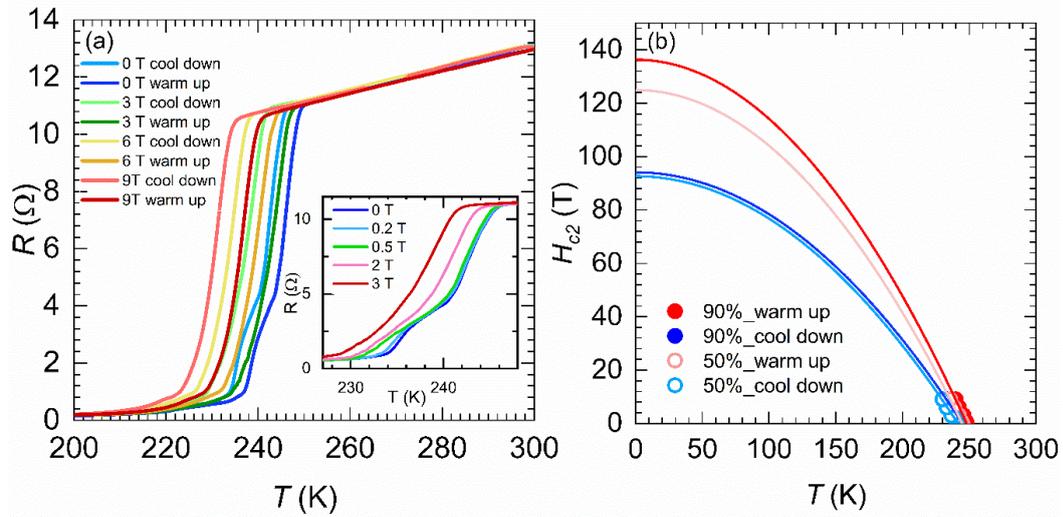

**Fig. 2| Superconducting transition under an external magnetic field.**
**a**, Electrical resistance $R$ as a function of the temperature $T$ for LaH$_{10}$ under applied magnetic fields up to $\mu_0 H$ = 9 T. The width of the superconducting transition remains essentially constant up to 9 T. Both, the cool down and the warm up temperature sweeps are plotted. The superconducting critical temperatures were determined as the average between both sweeps. An applied field of $\mu_0 H$= 9 T reduces the onset of the superconducting transition from ~250 K (as extracted from the warm up curve) to ~240 K. Notice that the step observed in the superconducting transition measured at zero-field, which appears around 245 K, disappears under the application of a modest field of just $\mu_0 H$ = 3 T. The gradual disappearance of this step via the application of an external magnetic field is shown in the inset. Likely, this step results from inhomogeneities in the superconducting sample. Similar behavior was found in disordered superconducting diamond doped with boron (Ref. 25).
**b**, Upper critical fields H$_{c2}$ as a function of the temperature following the criteria of 90% (solid markers) and 50% (open markers) of the resistance in the normal state. Solid curves are extrapolations resulting from fits to the Ginzburg-Landau expression in order to estimate the upper critical fields in the limit of zero temperatures. The temperature dependence of the observed upper critical fields $H_{c2}(T)$ were obtained from the $R(T,H)$ data displayed in Fig. 2(a). The $H_{c2}$s near $T_c$ increase nearly linearly with $T$ as $T$ is lowered. Here, we use a simple Ginzburg-Landau expression $H_{c2}(T) = H_{c2}(0)\left(1 - \left(T/T_c\right)^2\right)$ to estimate the upper critical fields $H_{c2}(0)$ in the limit of zero temperatures. The Ginzburg-Landau coherence lengths deduced from these $H_{c2}(0)$ values ranges between $\xi = 1.56$ and $1.86$ nm.

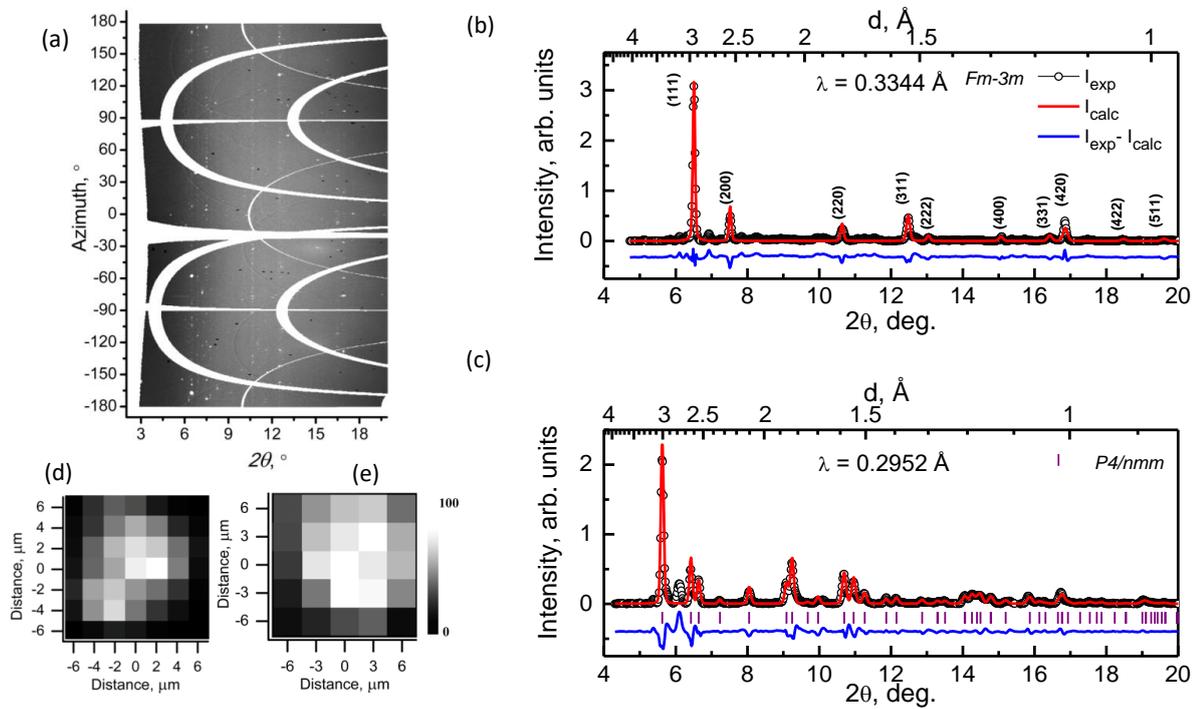

**Fig. 3| Structural analysis**. X-ray diffraction studies of superconducting lanthanum hydrides. (a) Cake representation of the typical X-ray powder diffraction pattern measured for the sample showing a superconducting step at ~249 K under 150 GPa (blue curve in Figure 1). The dominant face-centred *Fm-3m* phase gives a spotty powder pattern. Integrated powder patterns for samples synthesized from the La+H$_2$ mixture exhibiting a $T_c$ of ~249 K under 150 GPa (b) and the La+D$_2$ mixture with a $T_c$ ~140 K at 142 GPa (c). Indexation and refinement of these powder diffraction patterns yield two different phases, i.e. *Fm-3m* (*a* = 5.1019(5) Å, *V* = 132.80(4) Å$^3$) and *P4/nmm* (*a* = 3.7258(6) Å, *c* = 5.0953(12) Å and V = 70.73(2) Å$^3$) with stoichiometries corresponding to LaH$_{9.6}$ and LaD$_{10.6}$, respectively (see text and Extended Data for details). Both refined crystal structure models are in good agreement with the predicted structures for LaH$_{10}$ (Ref. 4,5) and LaH$_{11}$ (Ref. 3), except for the reflection at ~6.10° that cannot be indexed to the *P4/nmm* structural model (possibly it comes from an impurity phase). Black, red and blue graphs correspond to experimental data, fitted data, and the difference between experimental and fitted data. Le Bail method was chosen because it is impossible to extract the real intensities of the reflections from the spotty patterns. The distribution of these *Fm-3m* (d) and *P4/nmm* (e) phases throughout the samples is very smooth, containing more than 80% of the sample.

**Phases formed under hydrogen deficit**

We start by discussing the lowest hydride $LaH_3$, which is well studied at ambient and at low pressures. $LaH_3$ is an insulator and exhibits a pronounced Raman spectrum[1]. In our studies, we found that at $P > 100$ GPa, $LaH_3$ behaves as a poor metal or a semimetal, and it *does not exhibit SC* upon cooling down to ~5 K under pressures of at least 157 GPa.

Within the DAC, the lanthanum sample readily reacts with the surrounding hydrogen to form $LaH_3$, according to the Raman spectrum. This happens at $P$~10 GPa without any heating treatment or at room temperature. From the analysis of the X-ray powder diffraction (Extended Data Fig. 1) at the high pressure of 152 GPa, we determined the crystal structure of the sample to be $Fm3m$ with the refined lattice parameter $a = 4.3646(5)$ Å ($V = 83.14(3)$ Å$^3$, per formula unit $V = 20.78(3)$ Å$^3$). The $LaH_3$ stoichiometry for the sample was confirmed from the volume occupied by the three H atoms, that is ~5.7 Å$^3$ which is obtained after the extraction of the volume of the La atoms $V_{La}=15.1$ Å$^3$ (calculated using the equation of states (EOS) for La[2]) from the volume per formula unit V. The stoichiometry of the compound cannot be higher than 1:3 (e. g. $LaH_4$) since the volume per an H atom, $V_H$, is ~2 Å$^3$ according to extensive studies of hydrides under high pressures[3]. Lower stoichiometry (e.g. $LaH_2$) is also impossible according to Ref[4], because it is not stable at pressures above 11 GPa and separates into $LaH_3$ and $LaH_x$, where x is 0.25 or 0.6 - 0.7. Thus, the only stoichiometry that can correspond to the experimentally found cubic phase is $LaH_3$. From the $LaH_3$ stoichiometry we can accurately determine the volume taken by one H atom – which is ~1.9 Å$^3$ at 150 GPa. This estimation is valid even for higher pressures as the unit-cell parameter $a = 4.313(2)$ Å refined for the *fcc* $LaH_3$ phase found in the sample at 178 GPa gave the same $V_H$ ~1.9 Å$^3$. The obtained 1.9 Å$^3$ volume taken by one H atom at 150 GPa in $LaH_3$ gave us a key to estimate the stoichiometry of higher hydrides (or deuterides) as described in the main text.

In samples with an apparent hydrogen deficiency we found superconductivity but with lower $T_c$s. In particular, in the unheated mixture of La and $H_2$ pressurized up to 178 GPa, we observed a pronounced and reproducible superconducting step at $T_c$ ~70 K (Extended Data Fig. 2). The shift in $T_c$ to a lower temperature of ~49 K by an external magnetic field of 5 T further verifies the superconducting nature of this transition. By laser heating, the sample absorbed the rest of the hydrogen, its volume increased, and it transformed into a new superconducting phase with $T_c$ ~112 K (Extended Data Fig. 2). This suggests that increasing the hydrogen content (stoichiometry) would lead to an increase in $T_c$. This superconductivity was reproduced in another sample synthesized under 152 GPa, with almost the same value of $T_c$ of ~108 K. Notice that $T_c$ increases with pressure for this phase. X-ray diffraction patterns of the laser heated samples with $T_c$s of ~110 K and 70 K were found to be very complicated to analyze, perhaps because these samples are accompanied by other phases (Extended Data Fig. 1). According to theoretical predictions, the following stable metallic compounds could exist at $P$ ~150 GPa: $LaH_4$, $LaH_6$ and $LaH_8$,[5] (or only $LaH_6$[6]). The calculated $T_c$s are 5-10 K under 300 GPa, 150-160 K at 100 GPa and 114-150 K under 300 GPa[5], respectively. The experimentally found phase with $T_c$ ~110 K probably can be assigned to either $LaH_6$ or $LaH_8$.

**Phased formed under an excess of hydrogen**

Superconductivity with a much higher $T_c$ = 215 K was found in samples surrounded by a larger amount of hydrogen which seems to have been absorbed by the synthesis procedure (Extended Data Fig. 3); it is described in detail in Ref.[7]. Apparently, this phase competes with the phase exhibiting $T_c$ = 250 K as it can be synthesized under the same pressures following the aforementioned laser treatment procedure.

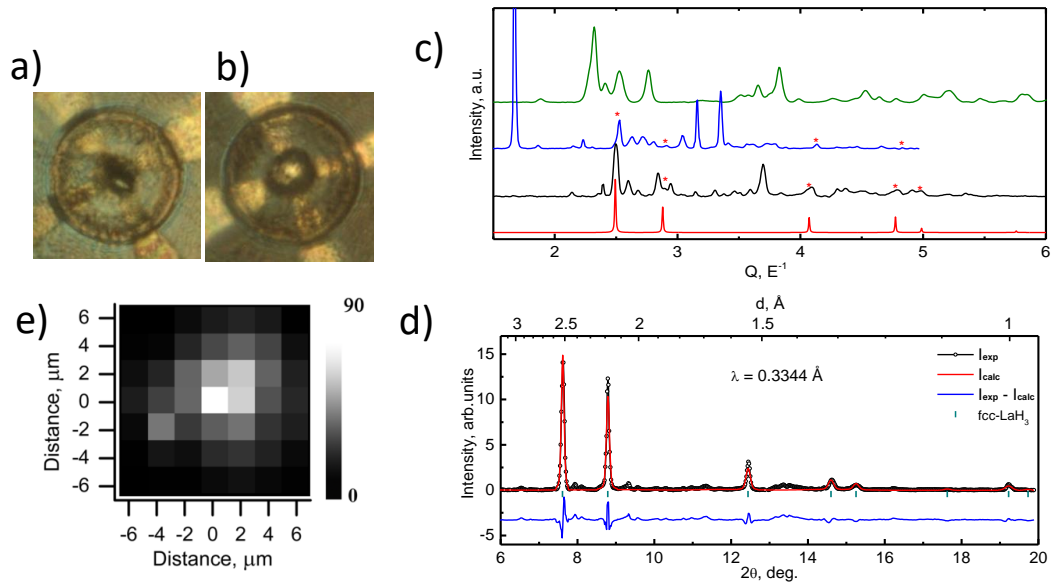

**Extended Data Fig. 1.** Characterization of the sample synthesized from a mixture of La+H$_2$ (in deficiency) at 150-180 GPa exhibiting superconductivity after laser heating. (a,b) Photographs of the sample exhibiting superconductivity with $T_c$ ~108 K before (a) and after (b) laser heating at ~1000 K at 152 GPa. Photos are taken in the combined transmitting-reflecting illumination. After thermal treatment, the sample size increased considerably and it started to reflect the incident light. (c) Typical integrated X-ray powder patterns for different samples synthesized from heating La in H$_2$ (in deficiency). The patterns seem like the ones from a mixture of different phases, and thus they cannot be indexed as a single phase. The black curve is for the sample synthesized at 152 GPa with $T_c$ ~108 K. Blue and green curves correspond to samples synthesized at 178 GPa ($T_c$ ~112 K) and 150 GPa ($T_c$ ~70 K), respectively. The red graph and red stars indicates the *Fm-3m* phase of LaH$_3$ (two of the samples contain some LaH$_3$ phase). (d) The integrated X-ray powder pattern of the sample shown in (b) measured in the center of the sample. It corresponds to almost pure *Fm-3m* phase of LaH$_3$. Black, red and blue graphs correspond to experimental data, fitted data, and the difference between experimental and fitted data. (e) The distribution of the *fcc* phase of LaH$_3$ in the heated sample, shown in (b), obtained from mapping the sample with X-ray focused beam. The brightest part in the center corresponds to the powder pattern presented in (d).

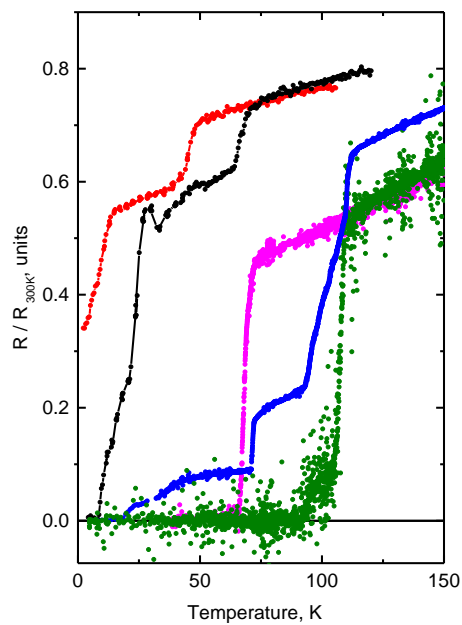

**Extended Data Fig. 2.** Superconductive transitions occurring in different samples prepared from a mixture of La+H$_2$, when there is H$_2$ deficiency. Resistance was normalized to the value at 300 K for each sample. The unheated mixture at 178 GPa, black curve, shows onset of the superconducting transition at ~70 K, which shifts with magnetic field to ~49 K at 5 T, red curve. The same superconductivity with $T_c$~70 K, magenta curve, was found in another sample prepared under 150 GPa by laser heating of La+H$_2$ mixture (in a great deficiency of H$_2$). After subsequent gradual laser heating of the first sample, black curve, up to ~1500 K, the sample absorbed the rest of the hydrogen, its volume increased, and a new superconductive transition appeared at ~112 K, blue curve. After several heating cycles, only one sharp transition at 112 K remained, green curve.

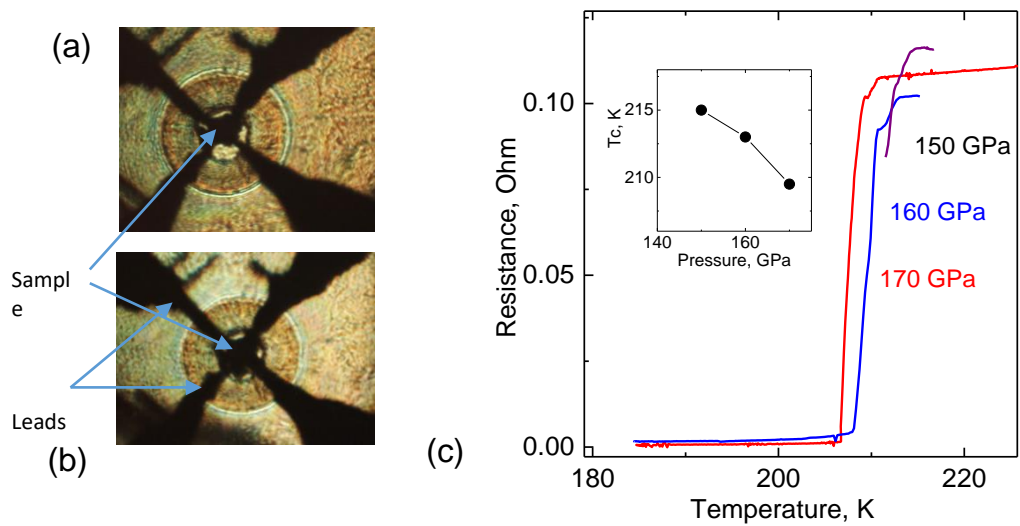

**Extended Data Fig. 3.** Superconductivity in a sample synthesized at 159 GPa from a mixture of La and an excess of $H_2$. View of the sample inside DAC with the attached four electrodes at transmission illumination (a) before and (b) after laser heating. As the result of heating, the sample significantly increased in volume and nearly filled the whole sample space, but the sample is still surrounded by hydrogen. (c) Superconducting steps at the temperature dependence of resistance at different pressures. The pressure dependence of the onset of $T_c$ is shown in the inset – $T_c$ shifts to lower temperatures as the pressure is increased.

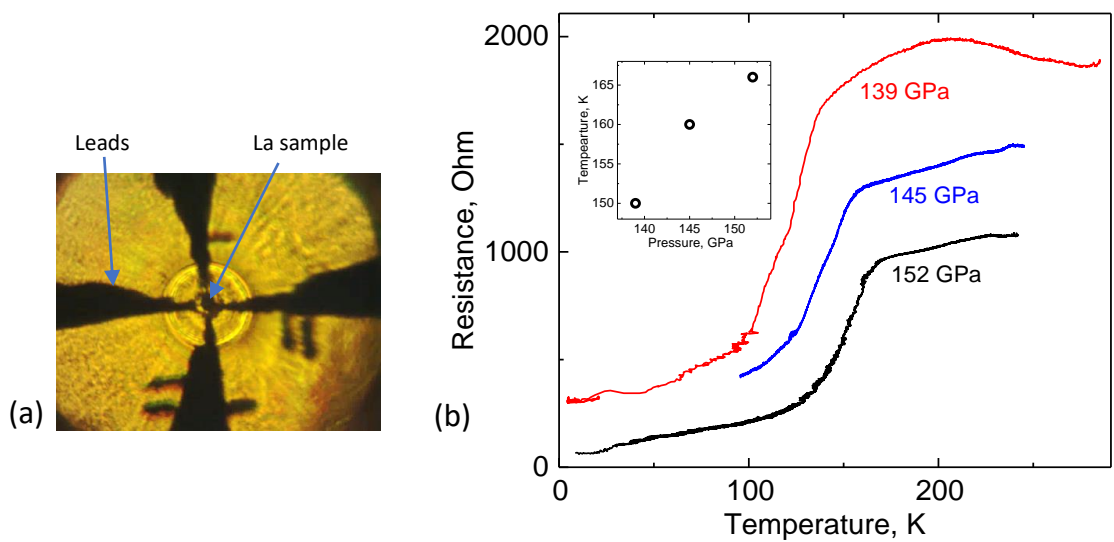

**Extended Data Fig. 4.** Lanthanum sample heated by laser in excess of deuterium. (a) Photograph of the heated La-D sample at 139 GPa taken in transmission illumination. (b) Superconducting transitions at different pressures (the pressure was determined from the $D_2$ vibron scale (Eremets, Troyan *Nat. Mat.*, **10**, 927–931 (2011)).

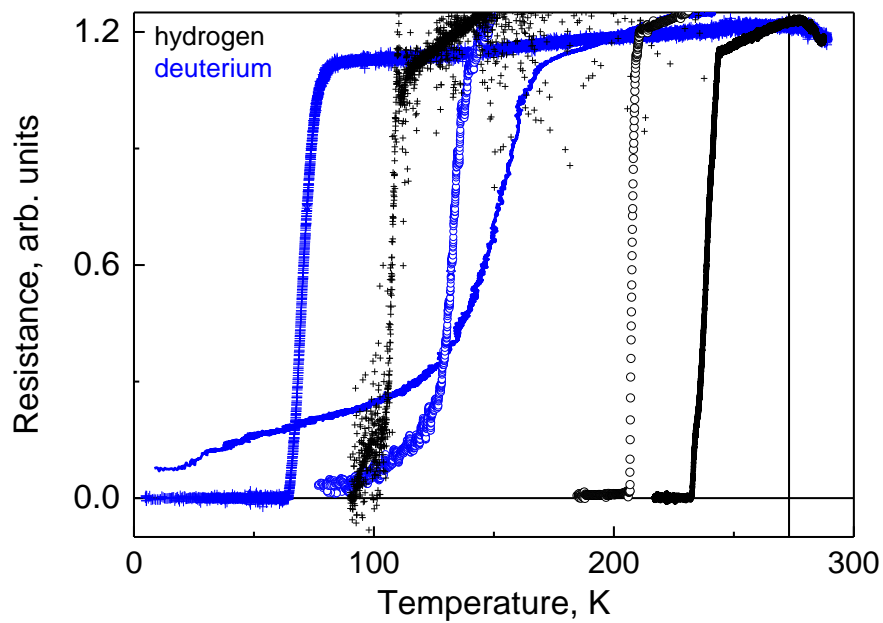

**Extended Data Fig. 5.** Superconducting transitions in hydrides synthesized from lanthanum in hydrogen (black plots) and deuterium (blue plots) atmospheres. The curves were scaled for a more direct comparison. Among the dozens of experiments performed, there are only four distinct, reproducible superconducting transitions in lanthanum hydrides with $T_c \sim$ 250 K, ~215 K, ~110 K, and three transitions in lanthanum deuterides with $T_c$ ~160 K, 140 K and ~70 K. For the hydrides series, we determined the structure only for the highest $T_c$ phase and obtained *Fm-3m* structure with the stoichiometry $LaH_{9.6}$ (Figure 3). For the deuterides series, we did not determine the structure for the phase with the highest $T_c$ ~160 K, but found that the next phase with $T_c$ ~140 K has different, tetragonal *P4/nmm* structure and $LaD_{10.6}$ stoichiometry (Figure 3).